# Ultrafast cryptography with indefinitely switchable optical nanoantennas

Pujuan Ma[1], Lei Gao[1], Pavel Ginzburg[2,3], and Roman E. Noskov[2,3,§]

[1]*School of Physical Science and Technology of Soochow University, Collaborative Innovation Center of Suzhou Nano Science and Technology, Soochow University, Suzhou 215006, China*

[2] *Department of Electrical Engineering, Tel Aviv University, Ramat Aviv, Tel Aviv 69978, Israel*

[3] *Light-Matter Interaction Centre, Tel Aviv University, Tel Aviv, 69978, Israel*

[§]*Correspondence: nanometa@gmail.com.*

**Abstract:** Bistability is widely exploited to demonstrate all-optical signal processing and light-based computing. The standard paradigm of switching between two steady states corresponding to "0" and "1" bits is based on the rule that a transition occurs when the signal pulse intensity overcomes the bistability threshold, and otherwise, the system remains in the initial state. Here, we break with this concept by revealing the phenomenon of *indefinite switching* in which the eventual steady state of a resonant bistable system is transformed into a nontrivial function of signal pulse parameters for moderately intense signal pulses. The essential nonlinearity of the indefinite switching allows realization of well-protected cryptographic algorithms with a single bistable element in contrast to software-assisted cryptographic protocols that require thousands of logic gates. As a proof of concept, we demonstrate stream deciphering of the word "enigma" by means of an indefinitely switchable optical nanoantenna. An extremely high bitrate ranging from ~0.1 to 1 terabits per second and a small size make such systems promising as basic elements for all-optical cryptographic architectures.

**Introduction**

Continuous growth in communication traffic drives new demands in novel cryptographic techniques that can supply reliable security of information exchange at a high rate. Many approaches have been actively proposed and explored. Although the majority of modern solutions focus on development of software algorithms[1,2], hardware realizations offer an additional and valuable degree of security. Recent examples of such approaches include digital metamaterials[3] and metasurfaces[4–6], which consider two elemental media with a strong contrast in the electromagnetic response as building blocks (treated as 'metamaterial bits'). The direct correspondence between the spatial structuring of such information bits and the electromagnetic properties of the entire system leads to a number of promising functionalities, such as information encoding in the radiation field patterns[5] as well as analogous[7] and digital[4] computations. Another actively developed realm involves securing optical information in images via linear and nonlinear holography[8,9]. However, all of these techniques typically require bulky structures and complicated programming algorithms[8,10,11], which restrict their performance.

Alternative strategies for all-optical information processing apply bistability. From a physical viewpoint, bistability occurs when a free-energy landscape of the system contains two minima separated by one maximum[12]. To switch between the minima, the system is disturbed with a form of activation energy, which is typically supplied by a large-amplitude input signal. For the all-optical bistable switchers reported thus far, this type of transition appears when the signal pulse intensity slightly overcomes the bistability threshold and the system returns to the initial state for weaker pulses[12–15]. Here, we show that this paradigm can be changed for sufficiently strong signal pulses that induce the activation energy to a much larger extent than the local maxima of the free energy. As a result, the eventual steady state is transformed to a nontrivial function of the pulse intensity, phase, and duration, as shown in Fig. 1(a). This phenomenon of *indefinite switching* is used to demonstrate cryptographically secure stream encryption.

Stream ciphers encrypt and decrypt bits individually using a secret key (a string of bits). To make ciphering sufficiently reliable, cryptographic devices must combine the randomness of a securing key and nonlinearity in the relationship between plaintext and ciphertext[1]. Importantly, vanishing nonlinearity results in a weak cryptographical security even when one has a random key with appropriate statistical properties because it is sufficient to know only a small portion of a message to fully unwrap the secret key by solving the linear system of equations (algebraic attack). Software-assisted cryptographic algorithms exploit combinations of many Boolean operations and shift registers to achieve nonlinearity. This approach calls for cascading of thousands of logic gates, which limits compactness and performance for such cryptographic architectures. For example, one of the fastest and most efficient modern stream ciphers, trivium, occupies approximately 4000 logic gates and encrypts at a maximal bitrate of 2 Gb s$^{-1}$ [1]. In contrast, indefinite switching naturally brings nonlinearity into the relationship between plaintext and ciphertext at the level of a single optical bistable nanoantenna. The ultrafast switching time and small size render such systems promising basic elements for novel lightweight cryptographic protocols operating at a Tb s$^{-1}$ bitrate.

In general, any side in a cryptographically secured data exchange process performs ciphering and deciphering procedures. Although these operations are the mathematical inverse of each other, they can be physically implemented by different devices. Here, we focus on deciphering by indefinite switching in an optical bistable nanoantenna (receiver), assuming data encoding in the intensities and phases of coherent signal pulses generated by an external laser system (sender).

As proof of concept, we consider scattering of a laser beam by a pair of nanoparticles (Fig. 1(b)). We assume that one nanoparticle in the dimer possesses a Kerr-type nonlinearity combined with a resonant response. In practice, such a nanoparticle can be constructed, for example, using any plasmonic material[16,17,18]. A surface plasmon resonance appears together with a strong cubic susceptibility for such materials as silver, gold and graphene[19–21]. This configuration allows driving of bistability at a moderate light power due to the strength of the nonlinear response reached via the resonance. The second nanoparticle is assumed to be linear and nonresonant and plays a role of a director to shape the directivity of the dimer scattering pattern.

Continuous-wave background radiation creates the desired nanoantenna energy landscape, giving rise to bistability, which together with the asymmetric dimer structure allows switching of the scattering pattern from an omnidirectional to a backward-scattered regime and back by illuminating the system with high-power signal pulses. Being easily distinguished in an experiment, such scattering states are associated with "0" and "1". With the aid of the general theoretical model, which describes the temporal transitions between these states, we show how to reach the regime of indefinite switching and elucidate its applications for ultrafast and efficient stream ciphering.

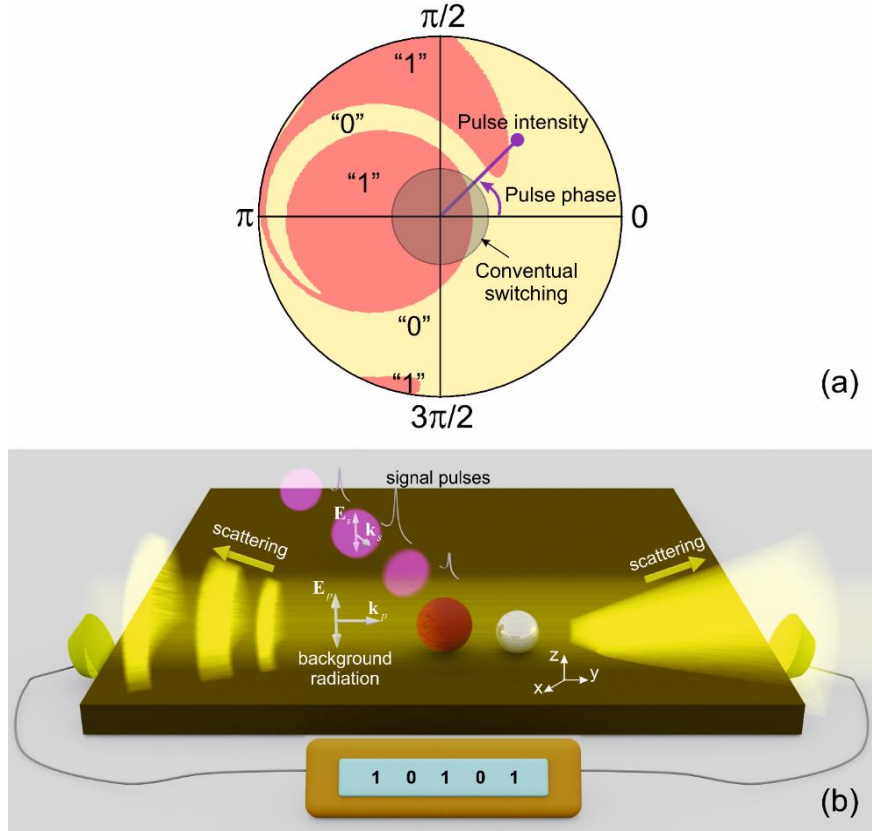

Fig. 1. General principle of indefinite switching and application for stream deciphering. (a) Switching diagram calculated by Eqs. 1 and 3. Polar angle and radii correspond to the signal pulse phase and intensity. Yellow and pink areas associated with "0" and "1" bits denote the eventual system state after disturbance by a signal pulse. The initial state is set to "1". The grey shadow area marks the regime of operation for conventional bistable switchers. A transition appears when the intensity of a signal pulse with an appropriate phase slightly overcomes the bistability threshold and the system remains in the initial position for weaker pulses, regardless of the pulse phase. The indefinite switching regime arises for stronger signal pulse intensities, transforming the eventual steady state into a nonlinear function of the pulse parameters. (b) Schematic demonstration of stream deciphering by indefinite switching implemented with a bistable heterodimer. Formed by continuous-wave background radiation, the heterodimer scattering pattern can take on omnidirectional or backward scattering states corresponding to "1" and "0" bits. Ciphertext, a sequence of signal pulses with distinct phases and intensities, is decoded into dual-bit plaintext by switching the nanoantenna scattering pattern, which is measured with aid of two synchronized photodiodes. The nonlinearity in the relationship between ciphertext and plaintext is supplied by indefinite switching, which warrants strong cryptographic security.

## Results

**Indefinite switching: Theoretical framework**

To perform a rigorous theoretical analysis of the system, the optical responses of subwavelength nanoparticles are treated within the point-dipole approximation. The dimer consists of a pair of particles in which one is resonant and nonlinear, and the second is nonresonant and linear (for the purpose of the directivity pattern switching). When the dimer is illuminated by a wave with the carrier frequency in the vicinity of the resonance ($\omega_0$), the first nanoparticle acts as a resonating oscillator with a slow (compared with the carrier) inertial response, whereas the time delay in the response of the second nanoparticle can be neglected because it is assumed to be nonresonant and dispersionless. With these considerations, we apply the dispersion relation method[22], which formulates the dynamic problem as a set of coupled equations for the slowly varying amplitudes of the nanoparticle dipole moments $P_1$ and $P_2$, written in dimensionless units as follows:

$$i\frac{dP_1}{d\tau} + \left(i\gamma + \Omega + |P_1|^2\right)P_1 = E\left(e^{ik_0 d} + \eta\right),$$
$$P_2 = \varsigma E + \eta P_1. \quad (1)$$

In this formulation, $E = E_p + E_s$ is the slow varying amplitude of the external light field containing pumping continuous-wave background radiation $E_p$ and signal pulse $E_s$ contributions, $\Omega = (\omega - \omega_0 - \delta\omega)/\omega_0$ is the driving frequency detuning from the resonant value, $\delta\omega$ is the resonance frequency shift due to the dipole–dipole coupling between the nanoparticles, $\gamma$ describes the thermal and radiation losses of the resonant nanoparticle, $k_0 = (\omega_0/c)\sqrt{\varepsilon_h}$, $c$ is the speed of light in vacuo, $\varepsilon_h$ is the permittivity of the host media, $\eta$ is the dipole-dipole coupling between nanoparticles, $\varsigma$ is the normalized director particle polarizability, and $\tau = \omega_0 t$ is the dimensionless time. The model derivation and detailed definitions of all these quantities are available in the Methods.

The steady-state solution of Eq. 1 is written as follows

$$\left(i\gamma + \Omega + |P_1|^2\right)P_1 = E\left(e^{ik_0 d} + \eta\right),$$
$$P_2 = \varsigma E + \eta P_1. \quad (2)$$

The nanoparticle polarizations $P_1$ and $P_2$ turn into three-valued functions of frequency with two stable branches and one unstable branch when $\Omega < -\sqrt{3}\gamma$. These branches feature different scattering patterns that can be found by substitution of Eq. 2 into the general expression for the normalized scattering pattern (far-field intensity) of a couple of dipoles[23]:

$$U(\varphi,\theta) = \sin^2\vartheta\left[|P_1|^2 + |P_2|^2 + 2|P_1||P_2|\cos(\Delta\kappa + k_0 d\sin\vartheta\sin\varphi)\right], \quad (3)$$

where $\varphi$ and $\theta$ are the spherical azimuthal and polar angles, and $\Delta\kappa$ denotes the phase shift between the dipoles. Because the director particle is nonresonant, efficient interparticle coupling occurs only when $k_0 d \ll 1$. In this case, the shape of the scattering pattern is defined predominantly by $\Delta\kappa$, altering the behavior from dipole-like for $\Delta\kappa \sim 0$ to pronounced directional shape for $\Delta\kappa \sim \pi$ [24]. Considering the negligible dispersion of the second nanoparticle, $\Delta\kappa$ is tunable mainly due to the resonance of the first nanoparticle. We also introduce the front-to-back scattering ratio as $U(\pi/2,\pi/2)/U(3\pi/2,\pi/2)$ to supply quantitative information on the dimer scattering properties.

**Indefinite switching: Realization**

The model represented by Eq. 1 is quite general and can be attributed to many different physical systems and engineering realizations. To make a qualitative estimate, we choose the following scenario: the heterodimer is realized as a pair of spherical nanoparticles made of graphene-wrapped ZnSe and silicon and placed in a free space. Graphene-based bistable switchers currently attract a great amount of attention[25–27]. The nanoparticle radii and the center-to-center distance are $R_{Si} = 120$ nm and $R_{ZnSe} = 50$ nm and $d = 400$ nm. We adjusted the parameters of the structure to obtain $\hbar\omega_0 = 0.133$ eV, which corresponds to the wavelength of a $CO_2$ laser at 9.32 μm (see further details in Methods).

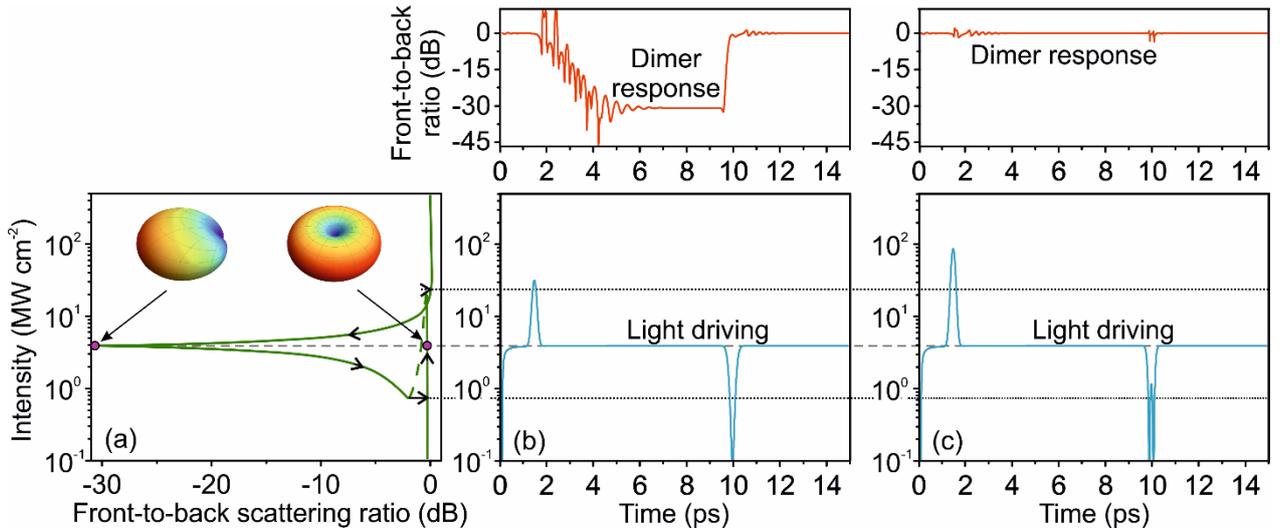

Fig. 2. Comparison of conventional bistability-induced switching and indefinite switching. (a) Steady-state front-to-back scattering ratio as a function of the background intensity given by Eqs. 2 and 3 with the detuning $\Omega = -0.12$. The bistability thresholds correspond to intensities 0.8 and 25 MW cm$^{-2}$ (indicated by dotted lines in (b) and (c)). The inserts show the scattering patterns at an intensity of 3.9 MWcm$^{-2}$ for which the steady states are marked by purple circles. Arrows denote the schematic transitions from one branch to another. The unstable branch is marked by the green dashed curve. (b) Demonstration of conventional temporal switching from the upper branch to the lower branch and back when the signal pulses slightly overcome the bistability thresholds. (c) The switching-off regime for signal pulses with peak intensities much stronger bistability thresholds. In panels (b) and (c), the driving light intensity and the resulted front-to-back scattering ratio are shown in the bottom and top rows, respectively. The pulse peak intensities and phases are (31.9 MW cm$^{-2}$, 0), (0.09 MW cm$^{-2}$, π) and (86.6 MW cm$^{-2}$, 0), (1.1 MW cm$^{-2}$, π) for (b) and (c), respectively. The pulse duration is 0.3 ps.

The typical bistable response of the heterodimer is shown in Fig. 2(a). Once the light intensity enters the bistability domain, the system demonstrates two stable solutions with omnidirectional dipole-like and pronounced backward scattering patterns, which we associate with information bits "1" and "0", respectively. The sharpest contrast in the front-to-back ratio between these states (~30 dB) appears for the intensity 3.9 MW cm$^{-2}$ and $\Delta\kappa = 0.1 \times \pi$ and $1.08 \times \pi$ for the "1" and "0" bits, in agreement with Eq. 3. We consider switching between these states, which, in practice, can be measured with a couple of synchronized photodiodes (Fig. 1(b)).

To induce transition from "0" to "1", following the common paradigm, one should illuminate the dimer with a signal pulse that is in phase with the background radiation and whose peak intensity together with the background intensity overcomes the high-intensity bistability threshold. An opposite transition from "1" to "0" is stimulated if one drives the system using an out-of-phase signal pulse with peak intensity, which is sufficiently large to decrease the total external field below the low-intensity bistability threshold. Figure 2(b) shows this type of switching, obtained by direct numerical simulation of Eq. 1. We assume that the background field slowly increases to the saturation level $E_0$ as $E_p = (2E_0/\pi)\arctan(\tau/\Delta\tau)$ ($\Delta\tau$ is the characteristic time of reaching the saturation), and the phase-locked signal pulse follows the Gaussian shape $E_s = E_{peak} \exp(-(\tau-\tau_0)^2/\delta^2 + i\Delta\Phi)$, where $E_{peak}$ is the peak amplitude of the pulse, $\tau_0$ defines the pulse temporal localization, $\delta$ is the pulse half-width, and $\Delta\Phi$ is the phase shift between the pulse and the background radiation. In practice, similar driving is widely used to realize coherent control over excitations in different nanostructures[28–30].

Surprisingly, we reveal that the system does not obey this rule because the peak pulse intensity grows considerably higher than the bistability thresholds. Specifically, the nanoantenna can transit to the counterpart state or remain in the initial position, depending on the signal pulse peak intensity, phase and duration. An example of such *indefinite switching* is shown in Fig. 2(c) when the switching is cancelled even though the signal peak intensities are significantly above the bistability thresholds.

Figure 3 shows the switching diagrams obtained by numerical simulations of Eq. 1. Specifically, we found the eventual heterodimer steady state by variations in the peak intensity, phase and duration of the signal pulse for the two initial states shown schematically in Figs. 3(a) and 3(b). When the pulse intensity remains comparatively low, the system obeys the conventional switching regime described above (Fig. 1(a)). However, as the intensity grows, the final state turns into a nontrivial function of the pulse characteristics due to an intricate interplay between the system inertia and nonlinearity and the stimuli, as shown in Figs. 3(c)-(j). Once the signal pulse passes, the system might be attracted by either the "0" or "1" state as a result of combined action of all of these factors. Additionally, the inertia yields the considerable increase of the switching thresholds with respect to the steady state prediction when the time of flight of a signal pulse decreases to below the decay time of oscillations, which is estimated as 0.3 ps (see Methods) (Figs. 3(g) and 3(h)).

The important characteristic of indefinite switching is the transitional time required to reach the eventual state, which also appears to be signal-pulse-dependent (Figs. 3(e), 3(f), 3(i), and 3(j)). Obviously, this time scale is minimal when the system remains in the initial position and increases once a transition occurs. These dependencies should be considered as the main limitation on the bitrate of information processing with the heterodimer because the temporal separation between the signal pulses must be longer than the system switching time to distinguish distinct bits.

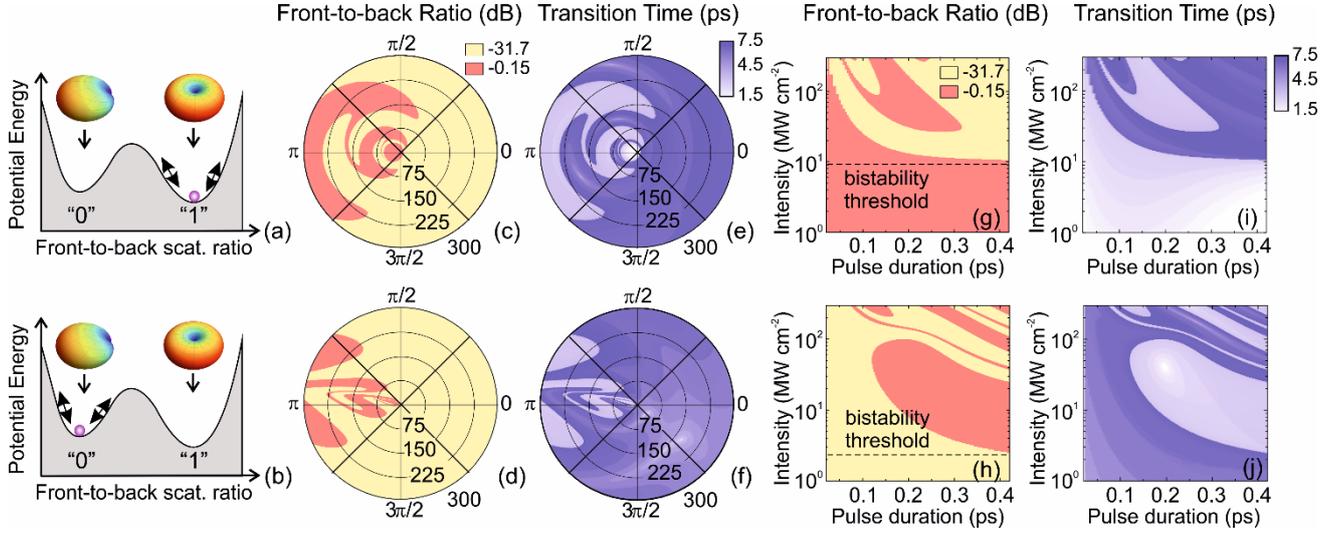

Fig. 3. Principle of indefinite switching implemented with a bistable heterodimer. (a), (b) Schematics of the system energy profile. Minima correspond to the stable steady states with omnidirectional and backward scattering patterns. The magenta circle indicates the system initial state. The results shown in the upper ((c), (e), (g), and (i)) and lower ((d), (f), (h), and (j)) rows were obtained for the initial states depicted in (a) and (b), respectively. In (c) and (d), the color denotes the front-to-back scattering ratio in the final state. The azimuthal angle and the polar radius denote the phase (with respect to the background radiation) and intensity (in MW cm$^{-2}$) of the signal pulse. In (e), (f), (i) and (j), the color shows the characteristic time required to reach the eventual steady state. The pulse duration for (c)-(f) is 0.4 ps. Panels (g)-(j) demonstrate the transitional diagrams in terms of signal pulse intensity and duration. For (g) and (i), the signal pulse is in phase with the background radiation, and for (h) and (j), it is out-of-phase. Other parameters are the same as in Fig. 2. The horizontal dashed lines in (g) and (h) mark the upper and lower steady-state bistability thresholds with respect to the background intensity.

**Decryption with indefinite switching**

An indefinitely switchable nanoantenna serves as a nonlinear function that converts incoming signal pulses with different intensities, phases, and durations into a binary output signal such that by relying on the switching diagrams in Fig. 3, one can build up a cipher, which allows conversion of a message hidden in a sequence of signal pulses (ciphertext) into a dual-bit format (plaintext). To demonstrate information decryption in this manner, we take 9 points belonging to the Archimedean spiral $\rho = \frac{300 \,[\mathrm{MW\,cm^{-2}}]}{3\pi} \Delta\Phi$ (the encryption key) at the phase-intensity switching map, as shown in Figs. 4(a) and 4(b), with the following parameters: 1 – (50 MW cm$^{-2}$, $\pi/2$), 2 – (75 MW cm$^{-2}$, $3\pi/4$), 3 – (150 MW cm$^{-2}$, $3\pi/2$), 4 – (175 MW cm$^{-2}$, $7\pi/4$), 5 – (200 MW cm$^{-2}$, 0), 6 – (225 MW cm$^{-2}$, $\pi/4$), 7 – (275 MW cm$^{-2}$, $3\pi/4$), 8 – (287 MW cm$^{-2}$, $7\pi/8$), and 9 – (300 MW cm$^{-2}$, $\pi$). Each of these pulses might drive switching to the counterpart state or may not, depending on the current dimer state. As the message, we choose the word "enigma", which in a dual-bit ASCII (American standard code for

information interchange) format is written as "01100101 01101110 01101001 01100111 01101101 01100001".

We launch a sequence of chosen signal pulses (Fig. 4(c)) with a fixed interval of 7.4 ps between them, which we adjusted based on Figs. 3(e) and 3(f) to avoid bit overlap. We stress that for successful decoding, one must appropriately set the background radiation intensity, the frequency detuning and the initial state. Figure 4(d) shows the resulting temporal response of the nanoantenna, which exactly reproduces the dual-bit code corresponding to the message. It is instructive to note that the bitrate of the demonstrated stream decryption is as fast as 240 light periods per one-bit processing or 0.13 Tb s$^{-1}$. This quantity is predominantly defined by the resonant frequency and the quality factor, which is ~30 for the considered system. Remarkably, the dipole plasmonic resonance in metallic nanoantennas is characterized by a similar quality factor. However, $\omega_0$ for such structures lies in the visible domain. Hence, for metallic-based nanoantennas, one can expect a bit rate of approximately several Tb s$^{-1}$.

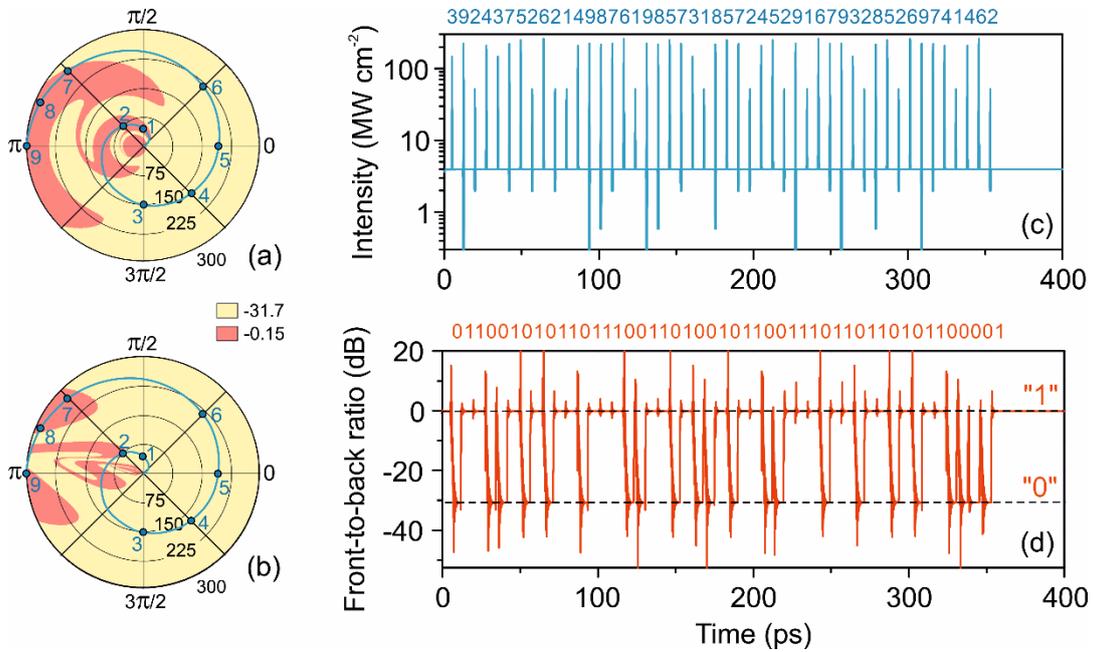

Fig. 4. Decryption of message encoded in the intensities and phases of signal pulses. (a), (b) Switching diagrams from Figs. 3(c) and 3(d) with the encryption key, i.e., the points belonging to the Archimedean spiral $\rho = \frac{300\ [\text{MW cm}^{-2}]}{3\pi}\Delta\Phi$, shown in blue. (c) Driving light field containing the background radiation and the sequence of ciphertext pulses, which are indicated in (a) and (b), as a function of time. (d) Resulting front-to-back scattering ratio for the excitation shown in (c). The eventual dual-bit code corresponds to the word "enigma".

## Discussion

The nonlinearity of indefinite-switching-assisted encryption means that a given plaintext bit can be encoded by a variety of ciphertext signal pulses, yielding an infinite number of ways to relate the ciphertext to the plaintext. Additional randomization of coding pulses renders the problem of identifying the correct coding key extremely challenging. It should be noted that this crucial requirement for reliability of cryptographic protocols is reached with only a single indefinitely switchable nanoantenna in contrast to software-assisted solutions, which are based on the use of thousands of logic gates.

The information exchange via indefinite switching relies on the sender's encoding algorithm (encryption key) and the sum of the receiver's system parameters (decryption key), which include the geometrical and material properties of the nanoantenna, the background radiation intensity, the frequency detuning and the initial state, and all of those parameters must be set prior to the deciphering procedure. Independent manipulations over these keys permit integration of indefinite switching and existing cryptographic principles.

In case of symmetric-key cryptography, all parties use the same encryption key and the same decryption key. Hence, the parties should have identical indefinitely switchable nanoantennas, and the keys should be shared in advance (before every encryption session) via a secured channel. Moreover, the background radiation intensity, frequency detuning and initial state play the role of the initialization vector (IV), which should take on a new random value for every encryption session. As the IV and the encryption key must be coherent, this procedure warrants that the encryption key must always be randomly different even if one encrypts the same message several times (semantic security)[1,2].

In asymmetric-key cryptography, the sides share only a public (encryption) key, which can be widely disseminated, and the private (decryption) key, which is different for every side, is held in secret. The pairing of these keys is based on the formalism of one-way functions, i.e., functions that are easy to compute on every input but computationally difficult to invert given the image of a random input. One-way functions are fundamental tools in personal identification, authentication, digital signatures and many other data security applications[1,2]. Indefinite switching can also operate as a one-way function since a given encoding algorithm can match many different decryption keys. Thus, once the public key is known, each party should use an individual nanoantenna with distinct properties to generate its own private key.

Importantly, the essential requirement for every cryptographic scheme is to maintain the carrier frequencies of the optical signals and the background radiation as equal. Unsynchronized light driving results in beating, which can be described as the following exchange in Eq. 1 $E_s \rightarrow E_s \exp(i\Delta\Psi(\tau)\tau)$, where $\Delta\Psi(\tau)$ is the frequency drift between the background radiation and a signal pulse in time. Obviously, this effect can lead to error switching unless the drift is much slower than the time duration of a decryption session $T$, i.e., $\Delta\Psi(\tau)T << \pi$. Another potential source of errors is nanoantenna damage owing to overheating, which might change its optical properties. However, a recent study[31] has shown that coating of plasmonic nanoantennas with a protective $Al_2O_3$ thin layer gives rise to unchanged both linear and nonlinear responses during 10-min illumination at a pulse repetition rate of 44 MHz and a pulse intensity 0.6 GW cm$^{-2}$. This observation allows us to believe that this issue is also avoidable.

To conclude, we have shown that a bistable optical heterodimer features indefinite switching, i.e., in response to a sufficiently strong signal pulse, the system can eventually transit to a counterpart steady state or remain in the initial position, depending on the pulse parameters. The essential nonlinearity of this operation makes indefinite-switching-enabled stream ciphering immune to algebraic attacks. In contrast to commonly used cryptographic algorithms requiring thousands of logic gates, indefinite-switching-assisted cryptographic protocols can be realized with a single bistable element, benefiting from compactness and performance. We demonstrated this principle by decrypting the word "enigma" with the remarkably high bit rate of 240 light periods per one-bit processing (or 0.13 Tb s$^{-1}$).

Our findings can be transferred to a variety of bistable systems, thus supplying great capacity for practical use. In particular, the presented general model is straightforwardly applicable to a broad range of nanoantennas, including plasmonic[32], core-shell[33–35] and high-index nanoparticles[36] as well as graphene flakes[37]. We believe indefinite switching can also be obtained in a wide range of other resonant bistable system, including multilayered structures[38,39], photonic crystal waveguides[15,40], and polaritonic cavities[41]. Many of these elements are readily available for integration in existing nanophotonic circuitry[14,42] and optical fibers[25,43], paving the way towards numerous opportunities for implementation of indefinite-switching-assisted cryptographic protocols on well-developed platforms. Additional highly important cryptographic functionalities, such as random secure key generation with appropriate statistical properties and integration of many bistable elements in a single device for parallelization, could enable engineering of all-optical cryptographic architectures operating at ultrafast speeds at the nanoscale.

## Methods

**General model derivation**

We assume that the dipole terms dominate in the nanoparticle responses and that quantum finite-size effects are negligible. The Kerr-like nonlinearity can be introduced via either the permittivity for nonlinear media as $\varepsilon_{NL} = \varepsilon_L + \Delta\varepsilon$ or the conductivity for graphene $\sigma_{NL} = \sigma_L + \Delta\sigma$. Here, nonlinear corrections $\Delta\varepsilon = \chi^{(3)}|\mathbf{E}_{in}|^2$ and $\Delta\sigma = \sigma^{(3)}|\mathbf{E}_{in}|^2$ contain the local electric field inside the nonlinear media $\mathbf{E}_{in}$, the cubic susceptibility $\chi^{(3)}$ and conductivity $\sigma^{(3)}$. The type of nonlinearity is assumed to be focusing since it was measured for the majority of highly nonlinear materials[19,20]. However, one can easily adjust our derivations for the defocusing nonlinearity as well.

We present the Fourier transforms of the nanoparticle electric-dipole moments as follows

$$\alpha_1^{-1} p_{1z} = E_z^{(ex)} \exp(ikd) + Gp_{2z},$$
$$p_{2z} = \alpha_2 \left( E_z^{(ex)} + Gp_{1z} \right),$$
(4)

where $E_z^{(ex)}$ is the driving electric field, and the nanoparticle polarizabilities $\alpha_1$ and $\alpha_2$ are given by the geometry of a specific system and analytically or numerically calculated or experimentally measured. The resonance frequency $\omega_0$ corresponds to the maximal nanoparticle polarizability $\alpha_1$. By appropriate fitting of the material composition, the nanoparticle shape and the host medium, one can tune this frequency from the visible range to the THz spectral domain. The electromagnetic Green's function, which describes the dipole-dipole interaction between the particles, is written as follows

$$G = \left[\frac{k^2}{d} + i\frac{k}{d^2} - \frac{1}{d^3}\right]\frac{e^{ikd}}{4\pi\varepsilon_0\varepsilon_h}$$

where $d$ is the center-to-center distance between the nanoparticles, $\varepsilon_0$ is the vacuum permittivity and $k = (\omega/c)\sqrt{\varepsilon_h}$ is the wavenumber in the host medium. Hereinafter, we assume harmonic dependence on time as $\exp[-i\omega t]$.

The dispersion relation method is based on the assumption that nonlinearity, losses, frequency detuning from resonance, and broadening of the particle polarization spectrum (due to temporal dynamics) are taken in the first order of perturbation, i.e., $\Delta\varepsilon \ll \varepsilon_L$, $\text{Im}\,\varepsilon_L \ll \text{Re}\,\varepsilon_L$, $(\omega-\omega_0)/\omega_0 \ll 1^{44}$. With this consideration, we decompose $\alpha_1^{-1}(\omega)$ in the vicinity of $\omega_0$,

$$\alpha_1^{-1}(\omega) \approx \partial_{\Delta\varepsilon}\alpha_1^{-1}\big|_{\substack{\text{Re}\,\varepsilon_L=0\\\omega=\omega_0}}\Delta\varepsilon + i\,\text{Im}\{\alpha_1^{-1}\}\big|_{\substack{\Delta\varepsilon=0\\\omega=\omega_0}} + \partial_\omega\alpha_1^{-1}\big|_{\substack{\Delta\varepsilon=0\\\text{Re}\,\varepsilon_L=0\\\omega=\omega_0}}\left(\Delta\omega + i\frac{d}{dt}\right). \tag{5}$$

In this formulation, we used the fact that $\text{Re}\{\alpha_1^{-1}\} = 0$ at resonance in a lossless case. We disregard the weak frequency-induced variations in the director polarizability $\alpha_2$ and the dipole coupling field and accept $\alpha_2 = \alpha_2(\omega_0)$ and $G = G(\omega_0)$.

Next, we assume that the nonlinear correction $\Delta\varepsilon$ does not change the structure of the particle eigenmode and results in the field-self action effect only[45] such that one can replace the spatially dependent local field in the graphene layer $\mathbf{E}_{in}$ with its average value

$$\langle|\mathbf{E}_{in}|^2\rangle = \frac{1}{V}\iiint_V |\mathbf{E}_{in}|^2 dv,$$

where $V$ is the volume of the nonlinear media. We express the nonlinear correction for the permittivity as follows

$$\Delta\varepsilon = \chi^{(3)}\psi\alpha_1^{-2}(\omega_0)|\mathbf{p}_{1z}|^2, \tag{6}$$

where $\psi = \langle|\mathbf{E}_{in}|^2\rangle/|\mathbf{E}^{(ex)}|^2$ is the field enhancement factor at resonance.

Using these considerations, we substitute Eqs. 5 and 6 into Eq. 4 and derive Eq. 1. The normalized quantities are given by the following

$$P_{1,2} = p_{1z,2z} \left[ \partial_{\Delta\varepsilon} \alpha_1^{-1}(\omega_0) \left\{ \partial_\omega \alpha_1^{-1}(\omega_0)\omega_0 \right\}^{-1} \psi \alpha_1^{-2}(\omega_0) \chi^{(3)} \right]^{1/2},$$

$$E = E^{(ex)} \left[ \partial_{\Delta\varepsilon} \alpha_1^{-1}(\omega_0) \left\{ \partial_\omega \alpha_1^{-1}(\omega_0)\omega_0 \right\}^{-3} \psi \alpha_1^{-2}(\omega_0) \chi^{(3)} \right]^{1/2},$$

$$\gamma = \mathrm{Im}\left\{ \alpha_1^{-1}(\omega_0) - G^2(\omega_0)\alpha_2(\omega_0) \right\} \left[ \partial_\omega \alpha_1^{-1}(\omega_0)\omega_0 \right]^{-1},$$

$$\delta\omega = \mathrm{Re}\left\{ G^2(\omega_0)\alpha_2(\omega_0) \right\} \left[ \partial_\omega \alpha_1^{-1}(\omega_0) \right]^{-1},$$

$$\varsigma = \alpha_2(\omega_0)\partial_\omega \alpha_1^{-1}(\omega_0)\omega_0, \quad \eta = \alpha_2(\omega_0)G(\omega_0).$$

To obtain the model for a graphene-based nanoantenna, one should make the following exchange: $\varepsilon_L \to \sigma_L$, $\Delta\varepsilon \to \Delta\sigma$, and $\chi^{(3)} \to \sigma^{(3)}$.

## Specific details of the heterodimer containing silicon and graphene-wrapped ZnSe nanoparticles

We assume that the heterodimer is placed in air with permittivity $\varepsilon_h = 1$ and contains a couple of spherical nanoparticles made of silicon and graphene-wrapped ZnSe. The nanoparticle radii and the center-to-center distance are $R_1 = R_{ZnSe} = 50$ nm and $R_2 = R_{Si} = 120$ nm and $d = 400$ nm. Because the gap between nanoparticle surfaces exceeds $\min\{R_{Si}, R_{ZnSe}\}$ and the light wavelength is much larger than the nanoparticle sizes, one can neglect boundary, nonlocal, and quantum finite-size effects[46] and apply the point-dipole approximation[47]. In the wavelength domain of 2.5-10 μm, dispersion for both silicon and ZnSe is insignificant such that we take their permittivities as $\varepsilon_{Si} \approx 11.7$[48] and $\varepsilon_{ZnSe} \approx 5.7$[49]. The linear graphene conductivity can be expressed for photon energies smaller than the Fermi energy in terms of the Drude model as follows[20]

$$\sigma_L = \frac{ie^2 E_F}{\pi\hbar^2 \left( \omega + i\xi^{-1} \right)},$$

where $e$ is the electron charge, $\hbar$ is the Planck constant, $\xi$ is the relaxation time for intraband transitions, $E_F = \hbar V_F \sqrt{\pi n}$ is the Fermi energy, $n$ is the doping electron density, and $V_F \approx c/300$ is the Fermi velocity. For high-quality graphene samples, one can estimate $\xi = 0.3$ ps and $E_F = 0.6$ eV. Hereinafter, we consider intraband transitions only. The assumption $k_B T \ll E_F$ ($k_B$ is the Boltzmann constant, and $T$ is the absolute temperature) allows us to disregard both interband transitions and temperature effects. The cubic conductivity is written in the local quasiclassical approximation as follows[20,21]

$$\sigma^{(3)} = \frac{-i9e^4 v_F^2}{8\pi E_F \hbar^2 \omega^3}.$$

The nanoparticle polarizabilities are expressed as $\alpha_2 = \alpha_{Si} = 6\pi\varepsilon_0 i a_1 / k^3$ ($a_1$ is the electric dipole Mie scattering coefficient) and

$$\alpha_1 = \alpha_{ZnSe} = 4\pi\varepsilon_0\varepsilon_h \left( \frac{\varepsilon_{ZnSe} + 2\varepsilon_h + 2i\sigma_{NL}[\omega R_{ZnSe}\varepsilon_0]^{-1}}{R_{ZnSe}^3 \left(\varepsilon_{ZnSe} - \varepsilon_h + 2i\sigma_{NL}[\omega R_{ZnSe}\varepsilon_0]^{-1}\right)} - i\frac{2}{3}k^3 \right)^{-1} {}^{50}. \quad (7)$$

The size of the silicon nanoparticle was adjusted to avoid the appearance of Mie resonances in the infrared spectral range. In contrast, the graphene-wrapped ZnSe nanoparticle possesses the pronounced resonance associated with the excitation of the dipole-type plasmon in the graphene shell. The eigenfrequency for this mode can be found from Eq. 7 and written in the linear and lossless case as follows

$$\omega_0 = \sqrt{\frac{2e^2 E_F}{R_{ZnSe}\varepsilon_0 \pi \hbar^2 (\varepsilon_{ZnSe} + 2\varepsilon_h)}}.$$

We adjusted the parameters of the structure to obtain $\hbar\omega_0 = 0.133$ eV, which corresponds to the wavelength of a $CO_2$ laser of 9.32 μm.

We estimate other parameters for this configuration as follows: $\psi \approx 550$, $\gamma \approx 0.0082$, $\delta\omega/\omega_0 \approx -7.4\times 10^{-6}$, $\varsigma \approx -56$, $\eta \approx -0.02$.

## Acknowledgments

This work was supported by PAZY Foundation, Kamin Project, the National Natural Science Foundation of China (Grant No. 11374223), the National Science of Jiangsu Province (Grant No. BK20161210), the Qing Lan project, the "333" project (Grant No. BRA2015353), and PAPD of Jiangsu Higher Education Institutions and China Scholarship Council (CSC). REN acknowledges useful discussions with Prof. N. A. Gippius, I. I. Shishkin, and A. Machnev.

**Competing interests:** The authors declare no competing financial interests.

## References

1   Paar C, Pelzl J. *Understanding Cryptography*. Springer: Berlin, Heidelberg; 2010.
2   Ferguson N, Schneier B, Kohno T. *Cryptography Engineering : Design Principles and Practical Applications*. Wiley: New York; 2010.
3   Della GC, Engheta N. Digital metamaterials. *Nat Mater* 2014; **13**: 1115–1121.
4   Cui TJ, Qi MQ, Wan X, Zhao J, Cheng Q. Coding metamaterials, digital metamaterials and programmable metamaterials. *Light Sci Appl* 2014; **3**: e218.

5   Cui TJ, Liu S, Li LL. Information entropy of coding metasurface. *Light Sci Appl* 2016; **5**: e16172.

6   Cui TJ, Liu S, Zhang L. Information metamaterials and metasurfaces. *J Mater Chem C* 2017; **5**: 3644–3668.

7   Silva A, Monticone F, Castaldi G, Galdi V, Alù A *et al*. Performing mathematical operations with metamaterials. *Science* 2014; **343**: 160–163.

8   Li LL, Cui TJ, Ji W, Liu S, Ding J *et al.* Electromagnetic reprogrammable coding-metasurface holograms. *Nat Commun* 2017; **8**: 197.

9   Chen JW, Wang K, Long H, Han XH, Hu HB *et al.* Tungsten disulfide–gold nanohole hybrid metasurfaces for nonlinear metalenses in the visible region. *Nano Lett* 2018; **18**: 1344–1350.

10  Moccia M, Liu S, Wu RY, Castaldi G, Andreone A *et al*. Coding metasurfaces for diffuse scattering: scaling laws, bounds, and suboptimal design. *Adv Opt Mater* 2017; **5**: 1700455.

11  Wu HT, Liu S, Wan X, Zhang L, Wang D *et al.* Controlling energy radiations of electromagnetic waves via frequency coding metamaterials. *Adv Sci* 2017; **4**: 1700098.

12  Gibbs HM. *Optical Bistability, Controlling Light with Light*. Academic: Orlando; 1985. 133.

13  Rosanov NN. *Spatial Hysteresis and Optical Patterns*. Springer-Verlag: Berlin, Heidelberg; 2002.

14  Liu L, Kumar R, Huybrechts K, Spuesens T, Roelkens G *et al.* An ultra-small, low-power, all-optical flip-flop memory on a silicon chip. *Nat Photonics* 2010; **4**: 182–187.

15  Li ZY, Meng ZM. Polystyrene Kerr nonlinear photonic crystals for building ultrafast optical switching and logic devices. *J Mater Chem C* 2014; **2**: 783–800.

16  Ginzburg P, Krasavin AV, Wurtz GA, Zayats AV. Non-perturbative hydrodynamic model for multiple harmonics generation in metallic nanostructures. *ACS Photonics* 2015; **2**: 8-13.

17  Krasavin AV, Ginzburg P, Wurtz GA, Zayats AV. Nonlocality-driven supercontinuum white light generation in plasmonic nanostructures. *Nat Commun* 2016; **7**: 11497.

18  Chen JS, Krasavin A, Ginzburg P, Zayats AV, Pullerits T *et al.* Evidence of high-order nonlinearities in supercontinuum white-light generation from a gold nanofilm. *ACS Photonics* 2018; **5**: 1927-1932.

19  Palpant B. Third-order nonlinear optical response of metal nanoparticles. In: Papadopoulos MG, Sadlej AJ, Leszczynski J, editors. *Non-Linear Optical Properties of Matter*. Dordrecht: Springer; 2006. p461-508.

20  Mikhailov SA, Ziegler K. Nonlinear electromagnetic response of graphene: frequency multiplication and the self-consistent-field effects. *J Phys Condens Matter* 2008; **20**: 384204.

21  Peres NMR, Bludov YV., Santos JE, Jauho AP, Vasilevskiy MI. Optical bistability of graphene in the terahertz range. *Phys Rev B* 2014; **90**: 125425.

22  Noskov RE, Belov PA, Kivshar YS. Subwavelength modulational instability and plasmon oscillons in nanoparticle arrays. *Phys Rev Lett* 2012; **108**: 093901.

23  Balanis CA. *Antenna Theory: Analysis and Design*. New York: Wiley-Interscience; 2005.

24  Shegai T, Chen S, Miljković VD, Zengin G, Johansson P *et al.* A bimetallic nanoantenna for directional colour routing. *Nat Commun* 2011; **2**: 481.

25  Gan XT, Wang YD, Zhang FL, Zhao CY, Jiang BQ *et al*. Graphene-controlled fiber bragg grating and enabled optical bistability. *Opt Lett* 2016; **41**: 603-606.

26  Gu T, Petrone N, McMillan JF, Van Der Zande A, Yu M *et al.* Regenerative oscillation and four-wave mixing in graphene optoelectronics. *Nat Photonics* 2012; **6**: 554-559.

27  Zhang K, Huang Y, Miroshnichenko AE, Gao L. Tunable optical bistability and tristability in nonlinear graphene-wrapped nanospheres. *J Phys Chem C* 2017; **121**: 11804-11810.

28  Feurer T, Vaughan JC, Nelson, KA. Spatiotemporal coherent control of lattice vibrational waves. *Science* 2003, **299**: 374-377.

29  Brinks D, Stefani FD, Kulzer F, Hildner R, Taminiau TH *et al*. Visualizing and controlling vibrational wave packets of single molecules. *Nature* 2010; **465**: 905-908.

30  Koehler JR, Noskov RE, Sukhorukov AA, Novoa D, Russell PstJ. Coherent control of flexural vibrations in dual-nanoweb fibers using phase-modulated two-frequency light. *Phys Rev A 2*017; **96**: 063822.

31  Albrecht G, Ubl M, Kaiser S, Giessen H, Hentschel M. Comprehensive study of plasmonic materials in the visible and near-infrared: linear, refractory, and nonlinear optical properties. *ACS Photonics* 2018; **5**: 1058-1067.

32  Drachev VP, Buin AK, Nakotte H, Shalaev VM. Size dependent $\chi^{(3)}$ for conduction electrons in Ag nanoparticles. *Nano Lett* 2004; **4**: 1535-1539.

33  Neuendorf R, Quinten M, Kreibig U. Optical bistability of small heterogeneous clusters. *J Chem Phys* 1996; **104**: 6348–6354.

34  Argyropoulos C, Chen PY, Monticone F, D'Aguanno G, Alù A. Nonlinear plasmonic cloaks to realize giant all-optical scattering switching. *Phys Rev Lett* 2012; **108**: 263905.

35  Yu WJ, Ma PJ, Sun H, Gao L, Noskov RE. Optical tristability and ultrafast fano switching in nonlinear magnetoplasmonic nanoparticles. *Phys Rev B* 2018; **97**: 075436.

36  Makarov SV, Zalogina AS, Tajik M, Zuev DA, Rybin MV *et al*. Light-induced tuning and reconfiguration of nanophotonic structures. *Laser Photon Rev* 2017; **11**: 1700108.

37  Smirnova DA, Noskov RE, Smirnov LA, Kivshar YS. Dissipative plasmon solitons in graphene nanodisk arrays. *Phys Rev B* 2015; **91**: 075409.

38  Noskov RE, Zharov AA. Optical bistability of planar metal/dielectric nonlinear nanostructures. *Opto-Electronics Rev* 2006; **14**: 217–223.

39  Husakou A, Herrmann J. Steplike transmission of light through a metal-dielectric multilayer structure due to an intensity-dependent sign of the effective dielectric constant. *Phys Rev Lett* 2007; **99**: 127402.

40  Yanik MF, Fan SH, Soljačić M, Joannopoulos JD. All-optical transistor action with bistable switching in a photonic crystal cross-waveguide geometry. *Opt Lett* 2003; **28**: 2506-2508.

41  Amo A, Liew TCH, Adrados C, Houdré R, Giacobino E *et al*. Exciton–polariton spin switches. *Nat Photonics* 2010; **4**: 361-366.

42  Chen R, Ng KW, Ko WS, Parekh D, Lu FL *et al*. Nanophotonic integrated circuits from nanoresonators grown on silicon. *Nat Commun* 2014; **5**: 4325.

43  Xomalis A, Demirtzioglou I, Plum E, Jung Y, Nalla V *et al*. Fibre-optic metadevice for all-optical signal modulation based on coherent absorption. *Nat Commun* 2018; **9**: 182.

44  Whitham GB. *Linear and Nonlinear Waves*. Hoboken, NJ, USA: John Wiley & Sons, Inc.; 1999.

45  Gao L, Li ZY. Self-consistent formalism for a strongly nonlinear composite: comparison with variational approach. *Phys Lett A* 1996; **219**: 324-328.

46  Thongrattanasiri S, Manjavacas A, García de Abajo FJ. Quantum finite-size effects in graphene plasmons. *ACS Nano* 2012; **6**: 1766-1775.

47  Olivares I, Rojas R, Claro F. Surface modes of a pair of unequal spheres. *Phys Rev B* 1987; **35**: 2453-2455.

48  Chandler-Horowitz D, Amirtharaj PM. High-accuracy, midinfrared (450 cm$^{-1}$⩽ω⩽4000 cm$^{-1}$) refractive index values of silicon. *J Appl Phys* 2005; **97**: 123526.

49  Connolly J, diBenedetto B, Donadio R. Specifications of raytran material. *Proc SPIE* 1979; **0181**: 141–144.

50  Christensen T, Jauho AP, Wubs M, Mortensen NA. Localized plasmons in graphene-coated nanospheres. *Phys Rev B* 2015; **91**: 125414.